\documentclass[conference, a4paper]{IEEEtran}
\IEEEoverridecommandlockouts

\usepackage{cite}
\usepackage{algorithm}
\usepackage[noend]{algpseudocode}
\usepackage{amsmath,amssymb,amsfonts}
\usepackage[pdftex]{graphicx}
\usepackage{subcaption}
\usepackage{textcomp}
\usepackage{xcolor}
\usepackage{pifont}
\usepackage{multirow}
\usepackage{makecell}
\usepackage{array}

\begin{document}

\title{Semantic Communications for Speech Recognition}
\author{Zhenzi Weng\IEEEauthorrefmark{1}, Zhijin Qin\IEEEauthorrefmark{1}, and Geoffrey Ye Li\IEEEauthorrefmark{2}\\
\small \IEEEauthorrefmark{1} Queen Mary University of London, London, UK\\
\small \IEEEauthorrefmark{2} Imperial College London, London, UK\\
\small \{zhenzi.weng, z.qin\}@qmul.ac.uk, geoffrey.li@imperial.ac.uk
}

\maketitle

\begin{abstract}
The traditional communications transmit all the source data represented by bits, regardless of the content of source and the semantic information required by the receiver. However, in some applications, the receiver only needs part of the source data that represents critical semantic information, which prompts to transmit the application-related information, especially when bandwidth resources are limited. In this paper, we consider a semantic communication system for speech recognition by designing the transceiver as an end-to-end (E2E) system. Particularly, a deep learning (DL)-enabled semantic communication system, named DeepSC-SR, is developed to learn and extract text-related semantic features at the transmitter, which motivates the system to transmit much less than the source speech data without performance degradation. Moreover, in order to facilitate the proposed DeepSC-SR for dynamic channel environments, we investigate a robust model to cope with various channel environments without requiring retraining. The simulation results demonstrate that our proposed DeepSC-SR outperforms the traditional communication systems in terms of the speech recognition metrics, such as character-error-rate and word-error-rate, and is more robust to channel variations, especially in the low signal-to-noise (SNR) regime.

\end{abstract}


\IEEEpeerreviewmaketitle

\section{Introduction}
Due to the breakthroughs of deep learning (DL) in various fields, the intelligent communications have been investigated and considered as a cutting-edge technique to solve the bottlenecks of the traditional communication systems\cite{qin2019deep}. Particularly, the DL-enabled intelligent communications have achieved lots of successes in physical layer communications\cite{8214233,ye2017power} and wireless resource allocations\cite{sun2018learning,8633948}.

The communication systems utilizing DL techniques are typically designed to transmit digital bit sequences and optimized by minimizing the bit-error rate (BER) or symbol-error rate (SER), which achieves the first level communications according to the categorization by Shannon and Weaver\cite{weaver1953recent}. Inspired by potentially higher performance and efficiency of the second level communications, the DL-enabled semantic communications have been investigated recently and have shown great potentials to break though the existing technique problems in the traditional communications systems.

Semantic information is relevant to the transmission goal at the receiver, which could be either source massage recovery or more intelligent tasks. In the cases of intelligent tasks, the semantic information only contains the task-related features while the other irrelative features will not be extracted or transmitted. As a result, the transmission bandwidth is reduced significantly. Moreover, the transmitted features can be recovered via minimizing the semantic error instead of BER or SER. The mechanism behind this is that semantic information takes into account the meaning and veracity of source data because they can be both informative and factual\cite{carnap1952outline}, besides, the semantic data can be compressed to a proper size by employing a lossless method\cite{basu2014preserving}. However, it is very challenging to define the semantic information or to represent semantic features by a precise mathematical formula.

Inspired by the end-to-end (E2E) communication systems developed to address the challenges in traditional block-wise commutation systems\cite{o2017introduction,8985539}, different types of sources have been considered on E2E semantic communication systems. Particularly, an initial research on semantic communication systems for text information has been developed in\cite{guler2018semantic}, which directly mitigates the semantic error when achieving Nash equilibrium. However, such a text-based semantic communication system only measures the performance at the word level instead of the sentence level. Thus, a further investigation about semantic communications for text transmission, named DeepSC, has been carried out in\cite{9398576} to deal with the semantic error at the sentence level with various length. Moreover, a lite distributed semantic communication system for text transmission, named L-DeepSC, has been proposed in\cite{xie2020lite} to address the challenge of IoT to perform the intelligent tasks.

Recently, there are also investigations on semantic communications for other transmission contents, such as image and speech. A DL-enabled semantic communication system for image transmission, named JSCC, has been developed in\cite{bourtsoulatze2019deep}. Based on JSCC, an image transmission system, integrating channel output feedback, can improve image reconstruction\cite{kurka2020deepjscc}. Similar to text transmission, IoT applications for image transmission have been carried out. Particularly, a joint image transmission-recognition system has been developed in\cite{lee2019deep} to achieve high recognition accuracy. A deep joint source-channel coding architecture, name DeepJSCC, has been investigated in\cite{jankowski2020joint} to process image with low computation complexity. 

Regarding the semantic commutations for speech information, our previous work developed an attention mechanism-based semantic communication system to restore the source message, i.e., reconstruct the speech signals\cite{Weng2101:Semantic}. However, in this paper, we consider an intelligent task at the receiver to recover the text information of the input speech signals. Particularly, we propose a DL-enabled semantic communication system for speech recognition, named DeepSC-SR, by learning and extracting the text-related semantic features from the readable speech signals, then recovering the text transcription at the receiver. The main contributions of this article can be summarized as threefold:
\begin{itemize}
\item A novel semantic communication system for speech recognition, named DeepSC-SR, is first proposed, which designs joint semantic and channel coding as an integrated neural network (NN).

\item Particularly, the convolution neural network (CNN) is employed to compress the source data and recurrent neural network (RNN) is utilized for feature learning and extracting from the input speech signals. By doing so, DeepSC-SR can recover the text transcription at the receiver by only transmitting text-related features.  

\item Moreover, a trained model with high robustness to channel variations is obtained by training DeepSC-SR under a fixed channel condition, and then facilitating it with good performance when coping with different testing channel environments.

\end{itemize}

The rest of this article is structured as follows. Section \uppercase\expandafter{\romannumeral2} introduces the model of semantic communication system for speech recognition and performance metrics. In Section \uppercase\expandafter{\romannumeral3}, the details of the proposed DeepSC-SR is presented. Simulation results are discussed in Section \uppercase\expandafter{\romannumeral4} and Section \uppercase\expandafter{\romannumeral5} draws conclusions.

\emph{Notation}: The single boldface letters are used to represent vectors or matrices and single plain capital letters denote integers. Given a vector $\boldsymbol x$, $x_i$ indicates its $i$th component, $\left\|\boldsymbol x\right\|$ denotes its Euclidean norm. $\boldsymbol Y\in\mathfrak R^{M\times N}$ indicates that $\boldsymbol Y$ is a matrix with real values and its size is $M\times N$. Superscript swash letters refer the blocks in the system, e.g., $\mathcal T$ in $\boldsymbol\theta^{\mathcal T}$ represents the parameter at the transmitter. $\mathcal{CN}(\boldsymbol m,\;\boldsymbol V)$ denotes multivariate circular complex Gaussian distribution with mean vector $\boldsymbol m$ and co-variance matrix $\boldsymbol V$. Moreover, $\boldsymbol a\ast\boldsymbol b$ represents the convolution operation on the vectors $\boldsymbol a$ and $\boldsymbol b$.
\section{System Model}
The semantic communication system for speech recognition aims to transmit and recover the information-related semantic features. In this section, we introduce the details of the considered system model and the adopted performance metrics are presented.

\subsection{Input Spectrum and Transcription}
The original speech sample sequence is converted into a spectrum before feeding into the transmitter. Particularly, the input sample sequence, $\boldsymbol m$, is first divided into $N$ frames by slicing it into $N$ time-slices. Then, the $N$ frames are converted into a spectrum via a series of operations, i.e., Hamming window, fast Fourier transform (FFT), logarithm, and normalization. The spectrum includes $N$ vectors, $\boldsymbol s=\left[{\boldsymbol s}_1,\boldsymbol\;{\boldsymbol s}_2,\;\dots,\;{\boldsymbol s}_N\right]$, and each vector, ${\boldsymbol s}_n$, $n\in\left[1,\;2,\;\dots,\;N\right]$, represents the characteristics of its corresponding frame. 

Moreover, here we introduce the transcription of a single speech sample sequence as $\boldsymbol t=\left[t_1,\boldsymbol\;t_2,\;\dots,\;t_K\right]$, each token $t_k$, $k\in\left[1,\;2,\;\dots,\;K\right]$, represent a character in the alphabet or a word boundary. In this paper, the considered language is English, there are 26 characters in the alphabet, besides, we add $apostrophe$, $space$, and $blank$ as three word boundaries, then we have 29 tokens, i.e., $t_k\in\overline{\boldsymbol t}$, $\overline{\boldsymbol t}=\left[\mathrm a,\;\mathrm b,\;\mathrm c,\;\dots,\;\mathrm z,\;apostrophe,\;space,\;blank\right]$.

\subsection{Transmitter}
\begin{figure}[tbp]
\includegraphics[width=0.45\textwidth]{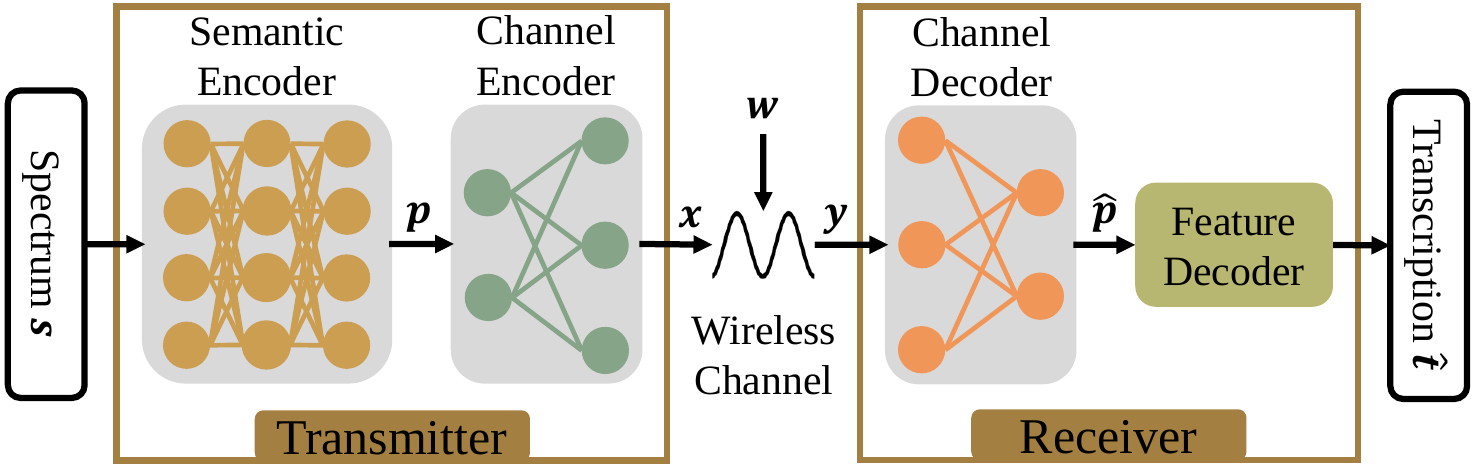} 
\centering 
\caption{Model structure of DL-enabled semantic communication system for speech recognition.}
\label{sys model}
\end{figure}

Based on the spectrum and transcription of the original speech sample sequence, the proposed system model is shown in Fig. \ref{sys model}. From the figure, the transmitter consists of two individual components: the \emph{semantic encoder} and the \emph{channel encoder}, each component is implemented by an independent NN. At the transmitter, the input spectrum is converted into the text features, $\boldsymbol p$, by the \emph{semantic encoder}, then the text features are mapped into symbols, $\boldsymbol x$, by the \emph{channel encoder} to be transmitted over physical channels. Denote the NN parameters of the \emph{semantic encoder} and the \emph{channel encoder} as $\boldsymbol\alpha$ and $\boldsymbol\beta$, respectively, the encoded symbol sequence, $\boldsymbol x$, can be expressed as
\begin{equation}
\boldsymbol x=\mathbf T_{\boldsymbol\beta}^{\mathcal C}(\mathbf T_{\boldsymbol\alpha}^{\mathcal S}(\boldsymbol s)),
\label{auto-encoder}
\end{equation}
where $\mathbf T_{\boldsymbol\alpha}^{\mathcal S}(\cdot)$ and $\mathbf T_{\boldsymbol\beta}^{\mathcal C}(\cdot)$ indicate the \emph{semantic encoder} and the \emph{channel encoder} with respect to (w.r.t.) parameters $\boldsymbol\alpha$ and $\boldsymbol\beta$, respectively. Here we denote the NN parameters of the transmitter as $\boldsymbol\theta^{\mathcal T}=(\boldsymbol\alpha,\boldsymbol\;\boldsymbol\beta)$.

The mapped symbols, $\boldsymbol x$, are transmitted over a physical channel. Note that the normalization on transmitted symbols $\boldsymbol x$ is required to ensure the transmission power constraint $\mathbb{E}\left\|\boldsymbol x\right\|^2=1$.

The whole transceiver in Fig. \ref{sys model} is designed for a single communication link, in which the wireless channel, represented by $p_h\left(\left.\boldsymbol y\right|\boldsymbol x\right)$, takes $\boldsymbol x$ as the input and produces the output as received signal, $\boldsymbol y$. Denote the coefficients of a linear channel as $\boldsymbol h$, then the transmission process from the transmitter to the receiver can be modeled as
\begin{equation}
\boldsymbol y=\boldsymbol h\ast\boldsymbol x+\boldsymbol w,
\label{channel}
\end{equation}
where $\boldsymbol w\sim\mathcal{CN}(0,\;\sigma^2\mathbf I)$ indicates independent and identically distributed (i.i.d.) Gaussian noise, $\sigma^2$ is noise variance for each channel and $\mathbf I$ is the identity matrix.

\subsection{Receiver}
The receiver also consists of two cascaded parts, including the \emph{channel decoder} and the \emph{feature decoder}, which commits to recover the text-related semantic features and convert them into the transcription as close as to the original transcription. Firstly, the received signal, $\boldsymbol y$, is mapped into the text-related semantic features, $\widehat{\boldsymbol p}$, by the NN-based \emph{channel decoder}. $\widehat{\boldsymbol p}=\left[{\widehat{\boldsymbol p}}_1,\;{\widehat{\boldsymbol p}}_2,\;\dots,\;{\widehat{\boldsymbol p}}_L\right]$ denotes a probability matrix, in which each probability vector ${\widehat{\boldsymbol p}}_l=\left[\widehat p_l^1,\;\widehat p_l^2,\;\dots,\;\widehat p_l^{29}\right]$, $l\in\left[1,\;2,\;\dots,\;L\right]$, consists of 29 probabilities and each probability represents that the $l$th token in ${\widehat{\boldsymbol p}}_l$ is the corresponding token in $\overline{\boldsymbol t}$. Denote the NN parameters of the \emph{channel decoder} as $\boldsymbol\delta$, then the recovered text features, $\widehat{\boldsymbol p}$, can be obtained from the received signal, $\boldsymbol y$, by
\begin{equation}
\widehat{\boldsymbol p}=\mathbf R_{\boldsymbol\delta}^{\mathcal S}(\boldsymbol y),
\label{auto-decoder}
\end{equation}
where $\mathbf R_{\boldsymbol\delta}^{\mathcal S}(\cdot)$ indicate the \emph{channel decoder} w.r.t. parameters $\boldsymbol\delta$. Denote the NN parameters of the receiver as $\boldsymbol\theta^{\mathcal R}=\boldsymbol\delta$.

Secondly, the text-related semantic features, $\widehat{\boldsymbol p}$, are decoded into the transcription, $\widehat{\boldsymbol t}$, by the \emph{feature decoder}, denoted as
\begin{equation}
\widehat{\boldsymbol t}=\mathbf R^{\mathcal F}(\widehat{\boldsymbol p}),
\label{feature decoder}
\end{equation}
where $\mathbf R^{\mathcal F}(\cdot)$ represents the \emph{feature decoder}.

Note that the process to map the probabilities into the transcription by the \emph{feature decoder} is unable to be implemented by the NN due to the non-differentiability. Moreover, the decoded transcription, $\widehat{\boldsymbol t}=\left[{\widehat t}_1,\boldsymbol\;{\widehat t}_2,\;\dots,\;{\widehat t}_{K'}\right]$, includes $K'$ tokens, which may be greater than, smaller than, or equal to $K$ tokens in the original transcription $\boldsymbol t$. 

The objective of the semantic communication system for speech recognition is to recover the text information of the original speech signals, which is equivalent to maximize the posterior probability $p\left(\left.\boldsymbol t\right|\boldsymbol s\right)$. Besides, by introducing connectionist temporal classification (CTC)\cite{graves2006connectionist}, the posterior probability $p\left(\left.\boldsymbol t\right|\boldsymbol s\right)$ can be expressed as
\begin{equation}
p\left(\left.\boldsymbol t\right|\boldsymbol s\right)=\sum_{A\in\mathfrak A(\boldsymbol s,\boldsymbol\;\boldsymbol t)}\left(\prod_{l=1}^L{\widehat p}_l\left(\left.a_l\right|\boldsymbol s\right)\right),
\label{posterior probability CTC}
\end{equation}
where $\mathfrak A(\boldsymbol s,\boldsymbol\;\boldsymbol t)$ represents the set of all possible valid alignments of transcription $\boldsymbol t$ to spectrum $\boldsymbol s$, and $a_l$ indicates the token under the valid alignments. For example, assume the transcription $\boldsymbol t=\left[\mathrm t,\boldsymbol\;\mathrm a,\;\mathrm s,\;\mathrm t,\;\mathrm e\right]$, the valid alignments could be $\left[blank,\;\mathrm t,\;blank,\boldsymbol\;\mathrm a,\;\mathrm s,\;blank,\;\mathrm t,\;\mathrm e\right]$, $\left[\mathrm t,\;blank,\boldsymbol\;\mathrm a,\;\mathrm s,\;blank,\;\mathrm{blank},\;\mathrm t,\;\mathrm e\right]$, etc. Note that the number of tokens in every valid alignment is $L$. Let a valid alignment be $\left[blank,\;\mathrm t,\;blank,\boldsymbol\;\mathrm a,\;\mathrm s,\;blank,\;\mathrm t,\;\mathrm e\right]$, the first token is $blank$, i.e., $a_1=blank$, then we have  
\begin{equation}
{\widehat p}_l\left(\left.a_l\right|\boldsymbol s\right)={\widehat p}_l\left(\left.blank\right|\boldsymbol s\right)=\widehat p_l^{29},\;\;\;\;\;l=1,
\label{probability pl}
\end{equation}
where $\widehat p_l^{29}$ is one of the probabilities in probability vector ${\widehat{\boldsymbol p}}_l$ and number 29 represents $blank$ is the 29th token in $\overline{\boldsymbol t}$.

Thus, for the sake of maximizing the posterior probability $p\left(\left.\boldsymbol t\right|\boldsymbol s\right)$, the CTC loss is adopted as the loss function in our system, which can be expressed as
\begin{equation}
{\mathcal L}_{CTC}(\boldsymbol\theta)=-\ln\left(\sum_{A\in\mathfrak A(\boldsymbol s,\boldsymbol\;\boldsymbol t)}\left(\prod_{l=1}^L{\widehat p}_l\left(\left.a_l\right|\boldsymbol s,\boldsymbol\theta\right)\right)\right),
\label{CTC loss}
\end{equation}
where $\boldsymbol\theta$ denotes the NN parameters of the transmitter and the receiver, $\boldsymbol\theta=(\boldsymbol\theta^{\mathcal T},\;\boldsymbol\theta^{\mathcal R})$. 

Moreover, given prior channel state information (CSI), the NN parameters, $\boldsymbol\theta$, can be updated by stochastic gradient descent (SGD) algorithm as follows:
\begin{equation}
\boldsymbol\theta^{(i+1)}\leftarrow\boldsymbol\theta^{(i)}-\eta\nabla_{\boldsymbol\theta^{(i)}}{\mathcal L}_{CTC}(\boldsymbol\theta),
\label{SGD}
\end{equation}
where $\eta>0$ is a learning rate and $\nabla$ indicates the differential operator.

\subsection{Performance Metrics}
In order to measure the similarity of the original and recovered text transcription, we adopt character-error-rate (CER) and word-error-rate (WER) in our model as the performance metrics. Based on the decoded transcription, $\widehat{\boldsymbol t}$, the substitution, deletion, and insertion operations are utilized to restore the original transcription, $\boldsymbol t$. Therefore, the calculation of CER can be denoted as
\begin{equation}
CER=\frac{S_C+D_C+I_C}{N_C},
\label{CER}
\end{equation}
where $S_C$, $D_C$, and $I_C$ represent the numbers of character substations, character deletions, and character insertions, respectively. $N_C$ indicates the number of characters in $\boldsymbol t$.  

Similarly, the WER can be calculated by
\begin{equation}
WER=\frac{S_W+D_W+I_W}{N_W},
\label{WER}
\end{equation}
where $S_W$, $D_W$, and $I_W$ denote the numbers of word substations, word deletions, and word insertions, respectively. $N_W$ represents the number of words in $\boldsymbol t$.

Note that the values of CER and WER may exceed 1 in some cases. Moreover, to evaluate the recognition accuracy for the same sentence, CER is typically lower than WER and it is generally acknowledged that the recognized sentence is readable when CER is lower than around 0.15.

\section{Semantic Communication System for Speech Recognition}
In this section, we present the details of the developed DeepSC-SR. Specifically, CNN and RNN are adopted for the semantic encoding, dense layer is employed for the channel encoding and decoding.

\subsection{Model Description}
\begin{figure*}[tbp]
\includegraphics[width=0.85\textwidth]{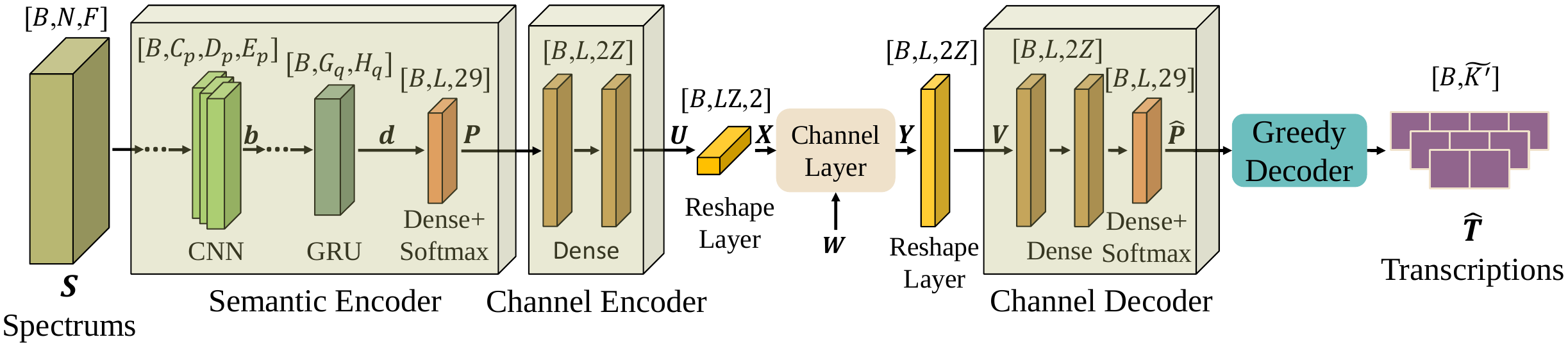} 
\centering 
\caption{The proposed system architecture for semantic communication system for speech recognition.}  
\label{proposed sys}
\end{figure*}

Let $\boldsymbol M$ be the set of speech sample sequences drawn from the speech dataset, which is converted in to a set of spectrums, $\boldsymbol S$. In addition, $\boldsymbol T\in\mathfrak R^{B\times\widetilde K}$ is the set of text transcriptions corresponding to $\boldsymbol M\in\mathfrak R^{B\times Q}$, where $B$ is the batch size, $Q$ is the number of samples in a sample sequence, and $\widetilde K$ indicates the number of tokens, $K$, in different batch of transcription is dynamic. The proposed DeepSC-SR is shown in Fig .\ref{proposed sys}. From the figure, the input of the proposed DeepSC-SR is spectrums, $\boldsymbol S\in\mathfrak R^{B\times N\times F}$, where $N$ denotes the number of frames and $F$ indicates the number of frequency bins corresponding to each speech frame. The spectrums, $\boldsymbol S$, are fed into the \emph{semantic encoder} to learn and extract the text-related semantic features and to output the features $\boldsymbol P\in\mathfrak R^{B\times L\times29}$. The details of the \emph{semantic encoder} are presented in part B of this section. Afterwards, the \emph{channel encoder}, denoted as two dense layers, converts $\boldsymbol P$ into $\boldsymbol U\in\mathfrak R^{B\times L\times2Z}$. In order to transmit $\boldsymbol U$ into a physical channel, it is reshaped into symbol sequences, $\boldsymbol X\in\mathfrak R^{B\times LZ\times2}$, via a reshape layer.

The received symbol sequences, $\boldsymbol Y$, are reshaped into $\boldsymbol V\in\mathfrak R^{B\times F\times2N}$ before feeding into the \emph{channel decoder}, represented by three dense layers. The output of the \emph{channel decoder} is the recovered text-related semantic features, $\widehat{\boldsymbol P}\in\mathfrak R^{B\times L\times29}$. Afterwards, the \emph{greedy decoder}, i.e., the \emph{feature decoder}, decodes $\widehat{\boldsymbol P}$ into the text transcriptions, $\widehat{\boldsymbol T}\in\mathfrak R^{B\times\widetilde{K^{'}}}$. Note that the number of tokens in different transcriptions is typically different. The details of the \emph{greedy decoder} are presented in part C of this section. The CTC loss is calculated at the end of the \emph{channel decoder} and backpropagated to the transmitter, thus, the trainable parameters in the whole system can be updated simultaneously.

\subsection{Semantic Encoder}\label{section semantic encoder}
The \emph{semantic encoder} is constructed by the CNN and the gated recurrent unit (GRU)-based bidirectional recurrent neural network (BRNN)\cite{650093} modules. As shown in Fig. \ref{proposed sys}, firstly, the input $\boldsymbol S$ is converted into the intermediate features via several CNN modules. Particularly, the number of \emph{filters} in each CNN module is $E_p$, $p\in\left[1,\;2,\;\dots,\;P\right]$, and the output of the last CNN module is $\boldsymbol b\in\mathfrak R^{B\times C_P\times D_P\times E_P}$. Secondly, $\boldsymbol b$ is fed into $Q$ BRNN modules, successively, and produces $\boldsymbol d\in\mathfrak R^{B\times G_Q\times H_Q}$, where the number of GRU units in each BRNN modules, $H_q$, $q\in\left[1,\;2,\;\dots,\;Q\right]$, is consistent. Finally, the text-related semantic features, $\boldsymbol P$, are obtained from $\boldsymbol d$ by passing through the cascaded dense layer and softmax layer.

\subsection{Greedy Decoder}\label{section Greedy Decoder}
\begin{figure}[tbp]
\includegraphics[width=0.45\textwidth]{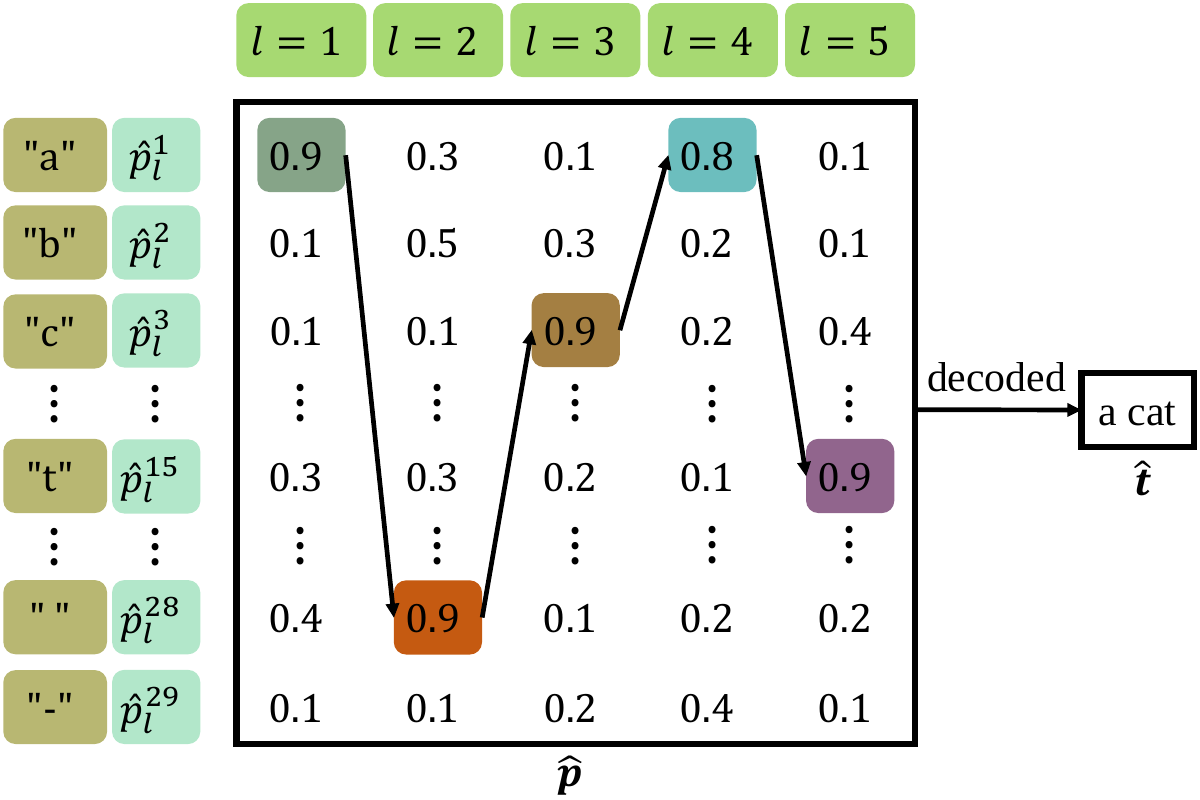} 
\centering 
\caption{An example of the \emph{greedy decoder}.}  
\label{greedy decoder}  
\end{figure}

As aforementioned that the recovered text features, $\widehat{\boldsymbol P}$, are decoded into the text transcriptions, $\widehat{\boldsymbol T}$, via the \emph{greedy decoder}. An example to obtain the transcription by the \emph{greedy decoder} is shown in Fig. \ref{greedy decoder}. During the decoding process, in each step $l$, the maximum probability in the probability vector, ${\widehat{\boldsymbol p}}_l$, is indexed and the token corresponding to this probability is employed to construct the final transcription $\widehat{\boldsymbol t}$.

\subsection{Model Training and Testing}
According to the prior knowledge of CSI, the training and testing algorithms of DeepSC-SR are described in Algorithm \ref{training algorithm} and Algorithm \ref{testing algorithm}, respectively. During the training stage, the \emph{greedy decoder} part is omitted. 
\begin{algorithm}[tbp]
\caption{Training algorithm of the proposed DeepSC-SR.}
\label{training algorithm}
\textbf{Initialization:} initialize parameters $\boldsymbol\theta^{(0)}$, $i=0$.

\begin{algorithmic}[1]
    \State \textbf{Input:} Speech sample sequences $\boldsymbol M$ and transcriptions $\boldsymbol T$ from trainset $\mathfrak S$, fading channel $\boldsymbol H$, noise $\boldsymbol W$.
    \State Generate spectrum $\boldsymbol S$ from sample sequences $\boldsymbol M$.
        \While{CTC loss ${\mathcal L}_{CTC}(\boldsymbol\theta)$ is not converged}
            \State $\mathbf T_{\boldsymbol\beta}^{\mathcal C}(\mathbf T_{\boldsymbol\alpha}^{\mathcal S}(\boldsymbol S))\rightarrow\boldsymbol X$.
            \State Transmit $\boldsymbol X$ and receive $\boldsymbol Y$ via (\ref{channel}).
            \State $\mathbf R_{\boldsymbol\delta}^{\mathcal S}(\boldsymbol Y)\rightarrow\widehat{\boldsymbol P}$.
            \State Compute loss ${\mathcal L}_{CTC}(\boldsymbol\theta)$ via (\ref{CTC loss}).
            \State Update parameters $\boldsymbol\theta$ via SGD according to (\ref{SGD}).
            \State $i\leftarrow i+1$.
        \EndWhile
    \State \textbf{end while}
    \State \textbf{Output:} Trained networks $\mathbf T_{\boldsymbol\alpha}^{\mathcal S}(\cdot)$, $\mathbf T_{\boldsymbol\beta}^{\mathcal C}(\cdot)$, and $\mathbf R_{\boldsymbol\delta}^{\mathcal C}(\cdot)$.
\end{algorithmic}

\end{algorithm}
\begin{algorithm}[tbp]
\caption{Testing algorithm of the proposed DeepSC-SR.}
\label{testing algorithm}

\begin{algorithmic}[1]   
    \State \textbf{Input:} Speech sample sequences $\boldsymbol M$ from testset, trained networks $\mathbf T_{\boldsymbol\alpha}^{\mathcal S}(\cdot)$, $\mathbf T_{\boldsymbol\beta}^{\mathcal C}(\cdot)$, and $\mathbf R_{\boldsymbol\delta}^{\mathcal C}(\cdot)$, testing channel set $\mathcal H$, a wide range of SNR regime.
    \State Generate spectrum $\boldsymbol S$ from sample sequences $\boldsymbol M$.
    	\For{channel condition $\boldsymbol H$ drawn from $\mathcal H$}
    	    \For{each SNR value}
    	        \State Generate Gaussian noise $\boldsymbol W$ under the SNR value.
    	        \State $\mathbf T_{\boldsymbol\beta}^{\mathcal C}(\mathbf T_{\boldsymbol\alpha}^{\mathcal S}(\boldsymbol S))\rightarrow\boldsymbol X$.
                \State Transmit $\boldsymbol X$ and receive $\boldsymbol Y$ via (\ref{channel}).
                \State $\mathbf R_{\boldsymbol\delta}^{\mathcal S}(\boldsymbol Y)\rightarrow\widehat{\boldsymbol P}$.
                \State Decoding $\widehat{\boldsymbol P}$ into $\widehat{\boldsymbol T}$ via (\ref{feature decoder}).
                \EndFor
            \State \textbf{end for}
        \EndFor
    \State \textbf{end for}
	\State \textbf{Output:} Recovered text transcriptions, $\widehat{\boldsymbol S}$.
\end{algorithmic}

\end{algorithm}

\section{Experiment and Numerical Results}
In this section, we compare to the performance between the proposed DeepSC-SR and the traditional communication systems under the AWGN channels and the Rayleigh channels, where the accurate CSI is assumed at the receiver. The traditional communication systems include \emph{benchmark 1} and \emph{benchmark 2}, which are introduced in the part A of this section. The experiment is conducted on the LibriSpeech dataset, which is a corpus of English speech with sampling rate of 16 kHz. Besides, the simulation environment adopted in our work is Tensorflow 2.3.

\subsection{Neural Network Setting and Benchmarks}
In the proposed DeepSC-S, the numbers of CNN modules and BRNN modules in the \emph{semantic encoder} are 2 and 6, respectively. For each CNN module, the number of \emph{filters} is 32, and for each BRNN module, the number of GRU units is 800. Moreover, two dense layers are utilized in the \emph{channel encoder} with units 40 and 40, respectively, and three dense layers are utilized in the \emph{channel decoder} with units 40, 40, and 29, respectively. The batch size is $B=16$ and learning rate is $\eta=0.0005$. The parameter settings of the proposed DeepSC-SR are summarized in Table \ref{DeepSC-SR NN parameters}. For performance comparison, we provide the following two benchmarks.

\subsubsection{\textbf{Benchmark 1}}
The first benchmark is a traditional communication system to transmit speech signals, named speech transceiver. Particularly, the input of the system is the speech signals, which is restored at the receiver. Moreover, the transcription is obtained from the recovered speech signals after passing through an automatic speech recognition (ASR) module. For the system, the adaptive multi-rate wideband (AMR-WB)\cite{1175533} is used for speech source coding and 64-QAM is utilized for modulation. Polar codes with successive cancellation list (SCL) decoding algorithm\cite{7114328} is employed for channel coding, in which the block length is 512 and the list size is 4. Moreover, the ASR module aims to recover the text transcript accurately, which is realized by employing Deep Speech 2\cite{pmlr-v48-amodei16} model.

\subsubsection{\textbf{Benchmark 2}}
The second benchmark is a traditional communication system to transmit text signals, named text transceiver. Particularly, the speech signals are converted into the text signals by the ASR model before feeding into the traditional system and the text transcription is recovered at the receiver. For the system, the Huffman coding\cite{4051119} is employed for text source coding, the settings of channel coding and modulation are same as that in \emph{benchmark 1}. In addition, Deep Speech 2 model is utilized to implement efficient speech recognition in the ASR module.
\renewcommand\arraystretch{1.15} 
\begin{table}[tbp]
\footnotesize
\caption{Parameter settings of the proposed DeepSC-SR.}
\label{DeepSC-SR NN parameters}
\centering
\begin{tabular}{|c|c|c|c|}
\hline
               & \textbf{Layer Name}  & \textbf{Filters/Units}  & \textbf{Activation}    \\
\hline
            \textbf{Semantic}        &   2$\times$CNN modules   & 2$\times$32   & ReLU \\
\cline{2-4}
            \textbf{Encoder}         &  6$\times$BCRNN modules  & 6$\times$800  & Tanh \\
\hline
            \textbf{Channel}        &        Dense layer        &    40         & ReLU \\
\cline{2-4}
            \textbf{Encoder}        &        Dense layer        &    40         & None \\
\hline            
   \multirow{3}{3.5em}{\textbf{\centering Channel\\Decoder}}  &  Dense layer  &  40  &  ReLU \\
\cline{2-4}
                                                              &  Dense layer  &  40  &  ReLU \\
\cline{2-4}
                                                              &  Dense layer  &  29  &  None \\
\hline
\end{tabular}
\end{table}

\subsection{Experiments}
\begin{figure}[tbp]
\includegraphics[width=0.41\textwidth]{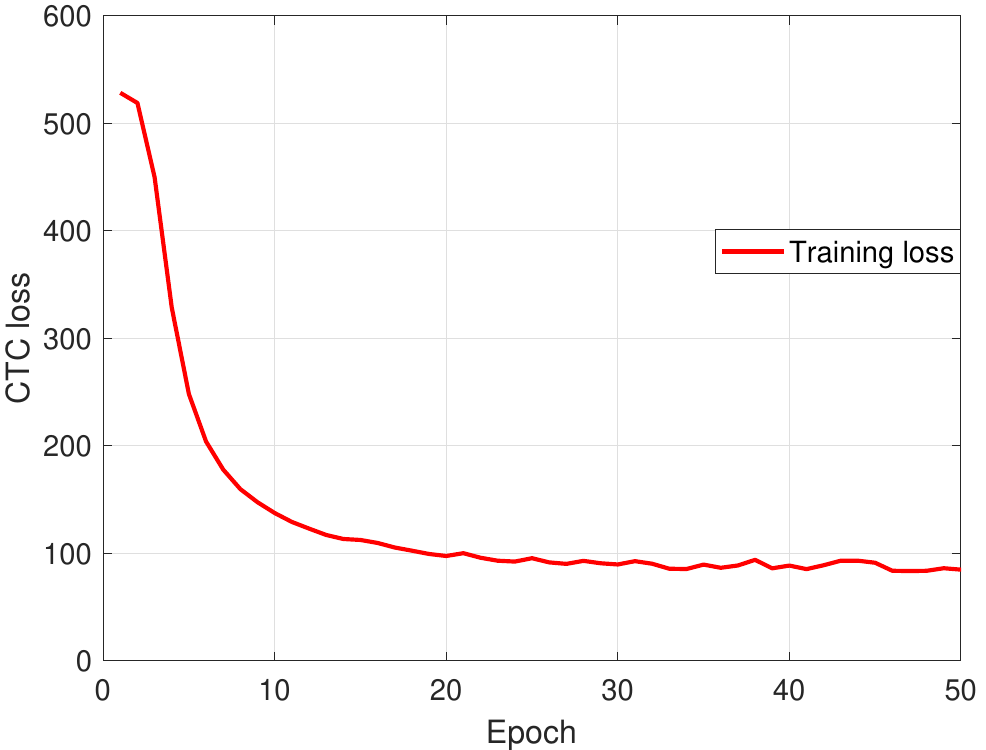}
\centering 
\caption{The training MSE loss versus epoch.}  
\label{loss vs epochs}  
\end{figure}

The relationship between the CTC loss and the number of epochs is shown in Fig. \ref{loss vs epochs}. From the figure, the CTC loss decreases to around 100 after 20 epochs and converges after about 50 epochs.

\begin{figure}[tbp]
\includegraphics[width=0.41\textwidth]{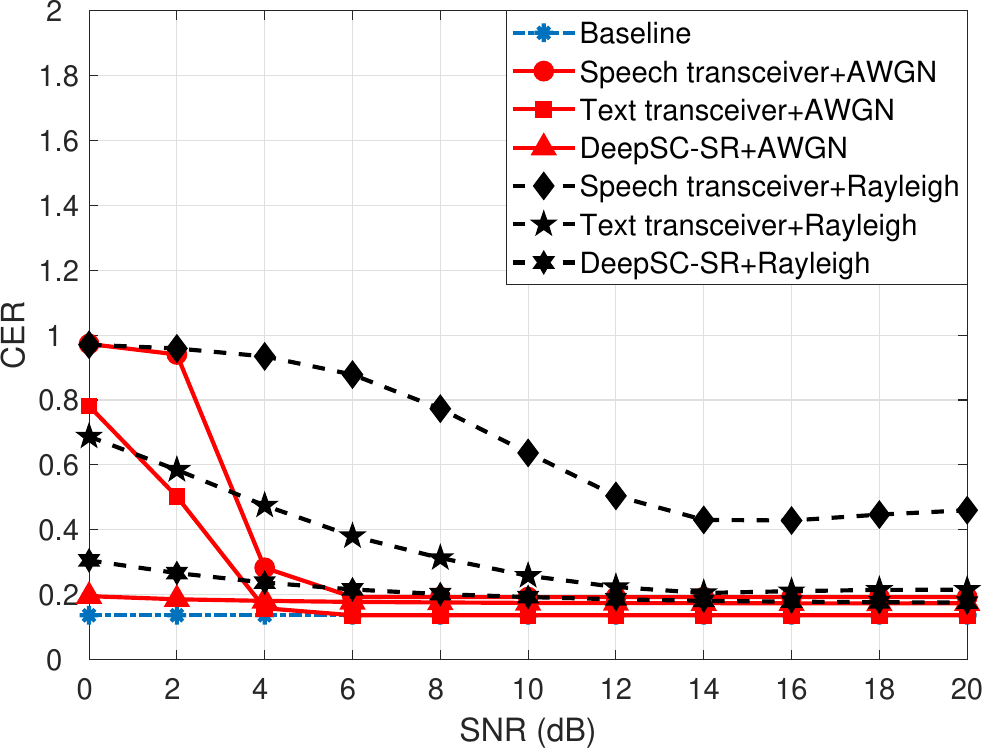}
\centering 
\caption{CER score versus SNR for the speech transceiver, the text transceiver, the proposed DeepSC-SR.}  
\label{CER result}  
\end{figure}

The CER results of DeepSC-SR and two benchmarks under the AWGN channels and the Rayleigh channels are shown in Fig. \ref{CER result}, where the baseline is the result tested by feeding the speech sample sequence into the ASR module directly without considering communication problems. From the figure, DeepSC-SR obtains lower CER scores than the speech transceiver and text transceiver under all tested channel environments. Moreover, DeepSC-SR performs steadily when coping with dynamic channels and SNRs while the performance of two benchmarks is quite poor under dynamic channel conditions. In addition, DeepSC-SR significantly outperforms the benchmarks in the low SNR regime. 

\begin{figure}[tbp]
\includegraphics[width=0.41\textwidth]{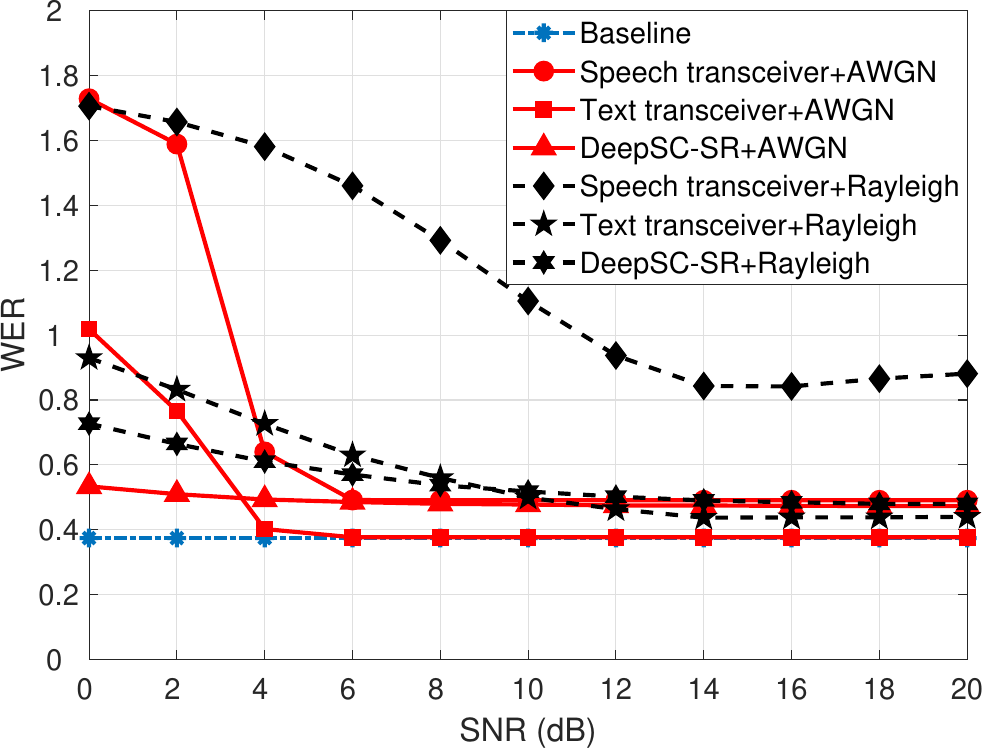}
\centering 
\caption{WER score versus SNR for the speech transceiver, the text transceiver, the proposed DeepSC-SR.}  
\label{WER result}  
\end{figure}

The WER scores comparison of different approaches are compared in Fig. \ref{WER result}. From the figure, the proposed DeepSC-SR can provide lower WER scores and outperform the speech transceiver under various channel conditions, as well as the text transceiver under the Rayleigh channels when SNR is lower than around 8 dB. Moreover, similar to the results of CER, DeepSC-SR obtains good WER scores on average when coping with channel variations while the traditional system provides poor scores when SNR is low. According to the simulation results, DeepSC-SR is able to yield better performance to recover text transcription at the receiver from the original speech signals at the transmitter when coping with the complicated communication scenarios than the traditional systems, especially in the low SNR regime.
\section{Conclusions}
In this article, we have investigated a DL-enabled semantic communication system for speech recognition, named DeepSC-SR, which aims to restore the text transcription by utilizing the text-related semantic features. Particularly, we jointly design the semantic and channel coding to learn and extract the features and mitigate the channel effects. Moreover, in order to facilitate DeepSC-SR adapting well over various physical channels, a model with strong robustness to channel variations has been investigated. Simulation results demonstrated that DeepSC-SR outperforms the traditional communication systems, especially in the low SNR regime. Hence, our proposed DeepSC-SR is a promising candidate for semantic communication systems for speech recognition.

\bibliographystyle{IEEEtran}
\bibliography{reference.bib}

\end{document}